\documentclass[conference]{IEEEtran}
\IEEEoverridecommandlockouts

\usepackage{cite}
\usepackage{amsmath,amssymb,amsfonts}
\usepackage{algorithmic}
\usepackage{graphicx}
\usepackage{textcomp}
\usepackage{xcolor}
\usepackage{booktabs}
\usepackage{multirow}

\def\BibTeX{{\rm B\kern-.05em{\sc i\kern-.025em b}\kern-.08em
    T\kern-.1667em\lower.7ex\hbox{E}\kern-.125emX}}

\begin{document}

\title{Learning Compact Terrain-Context Representations for Feasibility-Aware Offline Reinforcement Learning in UAV Relaying Networks}

\author{\IEEEauthorblockN{
Joseanne Viana \IEEEauthorrefmark{2}, 
Viswak R Balaji \IEEEauthorrefmark{3},
Boris Galkin \IEEEauthorrefmark{2}, 
Lester Ho \IEEEauthorrefmark{2},
Holger Claussen \IEEEauthorrefmark{2}\IEEEauthorrefmark{3}\IEEEauthorrefmark{4}
 }

\IEEEauthorblockA{\IEEEauthorrefmark{2} Tyndall National Institute, Dublin, Ireland;}
\IEEEauthorblockA{\IEEEauthorrefmark{3} University College Cork, Ireland;} \IEEEauthorblockA{\IEEEauthorrefmark{4} Trinity College Dublin, Ireland.}
Emails: 
\{joseanne.viana, lester.ho, Boris.Galkin, holger.claussen\}@tyndall.ie\vspace{-2em} 
}

\maketitle

\begin{abstract}
Offline reinforcement learning (RL) is an attractive tool for unmanned aerial vehicle (UAV) systems, where online exploration is costly and raises safety concerns. In terrain-aware UAV relaying, agents may observe high-dimensional inputs such as terrain and land-cover maps, which describe the propagation environment, but complicate offline learning from fixed datasets. This paper investigates the impact of compact state representations on offline RL for UAV relaying. End-to-end service is jointly constrained by UAV--user access links and a base-station--to--UAV backhaul link, yielding feasibility limits driven by user mobility and independent of UAV control. To distinguish feasibility limits from control-induced sub-optimality, a candidate-set feasibility upper bound (CS-FUB) is introduced, which estimates the maximum achievable user coverage over a restricted set of UAV placements. To address high-dimensional terrain context, map-like observations are compressed into low-dimensional latent representations using a variational autoencoder (VAE) and policies are trained via Conservative Q-Learning (CQL). Simulation results show that training CQL directly on raw high-dimensional terrain-context states leads to slow convergence and large feasibility gaps. In contrast, VAE-encoded representations improve learning stability, enable earlier convergence to feasible relay configurations, and  reduce sub-optimality relative to physical limits. Comparisons with autoencoder and linear compression baselines further demonstrate the benefit of structured representation learning for effective offline RL in terrain-aware UAV systems.
\end{abstract}

\begin{IEEEkeywords}
UAV communications, offline reinforcement learning, conservative Q-learning, variational autoencoder, terrain map, representation learning, sample efficiency
\end{IEEEkeywords}

\section{Introduction}

Unmanned aerial vehicles (UAVs) are increasingly deployed as \emph{aerial relays} \cite{Fonseca2023} to enhance the resilience and coverage of wireless networks, particularly in scenarios where terrestrial infrastructure is sparse, damaged, or temporarily unavailable. In such relay-based architectures, a UAV must simultaneously maintain a reliable \emph{backhaul link} to a ground base station and \emph{access links} to multiple mobile users. Consequently, feasible UAV placements and trajectories are determined by the joint satisfaction of access and backhaul constraints, which are strongly influenced by terrain, obstructions, and propagation effects. 

Reinforcement learning has been widely explored for UAV trajectory optimization \cite{Ho2024,Viana2024}, relay placement, and connectivity enhancement in wireless networks. Recent studies have applied RL to UAV-enabled relaying and cellular connectivity optimization, typically relying on online interaction and low-dimensional state representations~\cite{Zhang2023, Mehran2025}. At the same time, online RL is often impractical for real-world UAV relay systems due to safety constraints, limited interaction budgets, and deployment costs. Offline reinforcement learning~\cite{Levine2020, Kumar2020CQL, Kostrikov2022} has therefore emerged as an attractive alternative for wireless network control, a direction further emphasized by recent surveys in the communications literature~\cite{Eldeeb2024}. Despite its promise, incorporating terrain-aware coverage maps into offline RL poses significant challenges. Such map-like observations are inherently high-dimensional and exhibit strong spatial correlations. Naively incorporating coverage maps into reinforcement learning (RL) state representations substantially increases the complexity of value-function approximation, leading to poor sample efficiency and unstable learning behavior~\cite{Kumar2019Bootstrapping}.

In parallel, prior work on representation learning for control has shown that compact latent representations can improve learning stability and generalization~\cite{Zhang2021Invariant, Zhang2021Study, Lesort2018, Zeng2020}. Autoencoders and variational autoencoders (VAEs) provide a principled means of compressing high-dimensional observations while preserving structured information. This motivates a key question: how to represent terrain context efficiently.

This paper investigates whether compact representations enable stable and sample-efficient offline RL with terrain-aware inputs. To achieve this, high-dimensional terrain-derived coverage maps are encoded into low-dimensional latent states using a variational autoencoder, and conservative offline RL policies are trained on the resulting approximate latent decision process.

\subsection{Contributions}
This paper provides the following contributions:

\begin{itemize}
    \item An offline reinforcement learning formulation for \emph{UAV-assisted relaying} with joint access--backhaul constraints, in which the agent observes both low-dimensional kinematic (i.e ground user mobility) and link-quality features and a high-dimensional \emph{terrain-map}. 
    \item An investigation of \emph{compact latent representations} for terrain-aware relay control by encoding high-dimensional coverage maps into low-dimensional states using a variational autoencoder (VAE) and training conservative offline policies using Conservative Q-Learning (CQL).
    \item The introduction of a \emph{candidate-set feasibility upper bound} (CS-FUB) for feasibility-aware evaluation and reward normalization in offline reinforcement learning that estimates the maximum achievable end-to-end user coverage at each time step under joint access and backhaul constraints.
    \item Extensive simulation results comparing  VAE, autoencoder baselines, and linear compression via Principal Component Analysis (PCA) showing improved stability, faster convergence, and reduced feasibility gaps relative to learning from raw high-dimensional states.
\end{itemize}

\section{System Model and Problem Formulation}\label{sec:system}
A terrain-aware wireless relaying scenario is considered, consisting of a single UAV, a fixed ground base station, and $U$ mobile ground users.  The UAV operates as a relay and is responsible for positioning itself in three-dimensional space to provide wireless connectivity to users while maintaining a reliable connection to the base station. The operating region is described by a high-resolution digital terrain and vegetation cover map of Ireland, denoted by $M \in \mathbb{R}^{H \times W}$. The terrain component of the map consists of real-valued elevation measurements defined on an $H \times W$ grid, while the land-cover component encodes discrete surface classes (e.g., water, open land, sparse vegetation, and dense vegetation) that influence radio propagation through class-dependent attenuation \cite{NASA13}. Both components are spatially aligned and jointly characterize the physical environment \cite{NASA13}. At time $t$, the UAV is located at $x_t^{\mathrm{UAV}} \in \mathbb{R}^3$, while user
$u$ occupies position $x_{t,u}^{\mathrm{usr}} \in \mathbb{R}^3$. User positions
evolve according to pre-generated mobility trajectories and are not influenced
by UAV actions. The base station position
$x^{\mathrm{BS}} \in \mathbb{R}^3$ is fixed.

\subsection{MDP Formulation}
UAV control is modeled as a continuous-state Markov decision process (MDP) with a finite action set.
$\mathcal{M}=(\mathcal{S},\mathcal{A},P,r,\gamma)$ with discount factor $\gamma\in(0,1)$. At each time step $t$, the agent observes a state $s_t\in\mathcal{S}$, selects an action $a_t\in\mathcal{A}$, receives a reward
$r_t=r(s_t,a_t)$, and transitions to $s_{t+1}\sim P(\cdot\mid s_t,a_t)$. The objective is to learn a policy $\pi$ maximizing the expected discounted return. 
The transition dynamics and reward function implicitly encode the wireless feasibility constraints induced by the underlying system model, including both access and backhaul reliability. 

\subsection{Observation Structure: Terrain-Context State}
The agent observes both high-dimensional terrain context, and low-dimensional dynamic information, including the current UAV state, user coordinates, and received user power. The state at time $t$ is
\begin{equation}
s_t = (M,\, x_t,\, \ell_t),
\end{equation}
where $M\in\mathbb{R}^{H\times W}$ is a static digital elevation and land-cover
map, $x_t\in\mathbb{R}^{d_x}$ denotes UAV and user kinematics (i.e coordinates and speeds), and
$\ell_t\in\mathbb{R}^{d_\ell}$ contains link-quality indicators (i.e Received Signal Strength Indicator (RSSI)). The terrain map provides spatial context but does not directly encode link feasibility.

\subsection{Action Space and Motion Model}
The UAV selects motion actions $a_t \in [-1,1]^3$, corresponding to bounded displacements along $(x,y,z)$. In practice, actions are mapped to a discretized grid, yielding an effective discrete action set $\{-1,0,1\}^3$ used during dataset generation and training.

\subsection{Wireless Relaying and Feasibility}
At time $t$, a user is considered served if and only if: 
(i) the UAV--user access link satisfies a reliability threshold $\tau_a$, and
(ii) the base-station--to--UAV backhaul link satisfies a threshold $\tau_b$.
Let $\mathrm{AccessOK}(u,p,t)$ and $\mathrm{BackhaulOK}(p,t)$ denote these events.
The number of users served depends jointly on UAV placement and terrain-induced propagation effects. Due to user geometry and environmental constraints, there may exist time steps at which no UAV placement can satisfy all access and backhaul requirements simultaneously (i.e when users move too far apart).

\subsection{Candidate-Set Feasibility Upper Bound (CS-FUB)}
\begin{figure}[ht]
    \centering
    \includegraphics[width=\linewidth]{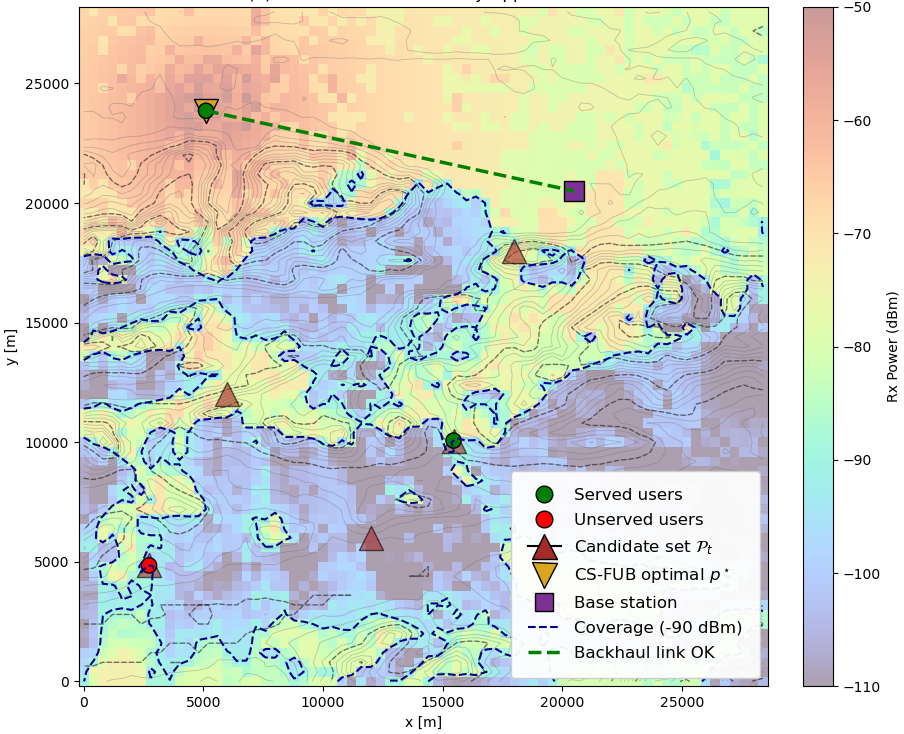}
\caption{%
CS-FUB illustration. Candidate UAV placements $\mathcal{P}_t$ and optimal location $p^\star$ are shown. Served and unserved users are indicated in green and red, respectively. The dashed blue contour denotes the coverage boundary (e.g., $-90$ dBm), highlighting geometry- and propagation-induced limits.
}
    \label{fig:cs_fub_map}
\end{figure}

Given the complexity of the environment, at a given time step there may be a fundamental upper limit on the achievable performance of the UAV relay, even under optimal placement. To represent this upper bound, the candidate-set feasibility upper bound (CS-FUB) is introduced. At time step $t$, let $\mathcal{P}_t \subset \mathbb{R}^3$ denote a finite set of candidate UAV placements. This candidate set represents feasible UAV positions for diagnostic evaluation only and is independent of the learned control policy (Fig.~\ref{fig:cs_fub_map}). In the experiments, $\mathcal{P}_t$ is instantiated as a finite set over the operational region (e.g., a discretized spatial grid with a small number of altitude levels) and evaluated offline. In addition, $\mathcal{P}_t$ includes user-centered positions (e.g., directly above users) and a centroid-based placement, providing a hybrid set that captures both global and local configurations.

For a given candidate placement $p \in \mathcal{P}_t$, the number of users that can be served at time $t$ is
\begin{equation}
N_{\mathrm{srv}}(p,t) =
\sum_{u=1}^{U}
\mathbf{1}\!\left[
\mathrm{AccessOK}(u,p,t)
\wedge
\mathrm{BackhaulOK}(p,t)
\right],
\end{equation}
where $\mathbf{1}[\cdot]$ denotes the indicator function, and $\mathrm{AccessOK}(u,p,t)$ and $\mathrm{BackhaulOK}(p,t)$ indicate whether the access and backhaul links satisfy their reliability thresholds.

The CS-FUB at time $t$ is then defined as
\begin{equation}
N^\star_{\mathrm{CS}}(t) =
\max_{p \in \mathcal{P}_t} N_{\mathrm{srv}}(p,t),
\end{equation}
which represents the maximum number of users that can be served by selecting the best placement from $\mathcal{P}_t$. The tightness of this bound depends on the density and diversity of $\mathcal{P}_t$, with finer discretization yielding a closer approximation to the continuous optimum.

In addition to evaluation, the CS-FUB is used during dataset generation for reward normalization. Rewards are scaled relative to $N^\star_{\mathrm{CS}}(t)$, allowing the learning process to distinguish feasibility-limited performance from control-induced suboptimality. Importantly, the CS-FUB is not used to guide action selection or generate trajectories.

\subsection{Offline Dataset}
Access to a fixed offline dataset is assumed
\begin{equation}
\mathcal{D}=\{(s_i,a_i,r_i,s'_i)\}_{i=1}^N,
\end{equation}
collected using a mixture of heterogeneous behavior policies, including random motion, centroid tracking, and coverage-oriented heuristics. The resulting data distribution is highly multi-modal and contains both sub-optimal and feasible behaviors. Each transition is generated by interacting with the environment. States capture the terrain and network configuration, actions represent UAV motion, and rewards reflect coverage and feasibility objectives. All learning is performed offline, without environment interaction.

\section{Methodology: VAE-Encoded Offline Reinforcement Learning}\label{sec:method}

\subsection{Terrain-Context Compression via VAE}
The terrain-context observations are compressed into compact latent representations using a variational autoencoder (VAE)~\cite{Lesort2018,Zhang2021Study}. Let $o_t \in \mathbb{R}^{d_o}$ denote the vectorized observation constructed from $(M,x_t,\ell_t)$.
The encoder maps $o_t$ to a latent distribution
\begin{equation}
(\mu_t,\log\sigma_t^2)=f_\phi(o_t),
\end{equation}
from which a latent sample is obtained via
\begin{equation}
z_t=\mu_t+\sigma_t\odot\epsilon,\quad \epsilon\sim\mathcal{N}(0,I).
\end{equation}
The decoder reconstructs $\hat{o}_t=g_\psi(z_t)$. The VAE is trained by minimizing
\begin{equation}
\small
\mathcal{L}_{\text{VAE}}=
\mathbb{E}\!\left[\|o_t-\hat{o}_t\|_2^2\right]
+\beta_{\text{KL}}D_{\text{KL}}\!\left(
\mathcal{N}(\mu_t,\sigma_t^2I)\,\|\,\mathcal{N}(0,I)\right).
\end{equation}
where $\phi$ and $\psi$ denote encoder and decoder parameters, and $\beta_{\text{KL}}$ weights the  Kullback-Leibler (KL) regularization. In this context, the VAE compresses terrain-induced propagation patterns and spatial correlations relevant to link feasibility. The latent representation captures features such as elevation and land-cover attenuation, enabling a compact state that preserves feasibility-relevant information. This regularization constrains the latent distribution, reducing overfitting and improving generalization under distributional shift in offline RL~\cite{Lesort2018}. The latent dimension $d_z$ balances compression and information preservation; moderate values (e.g., $d_z=64$) provide a good trade-off between stability and performance.

\subsection{Offline Control with Conservative Q-Learning}
Policies are trained using Conservative Q-Learning (CQL)~\cite{Kumar2020CQL}, which mitigates overestimation under distributional shift.

\subsection{Training Pipeline}
Training proceeds in two stages:
\begin{enumerate}
    \item The VAE is trained on observations from the offline dataset.
    \item The dataset is encoded into latent states, and CQL is trained on the resulting latent MDP.
\end{enumerate}

\section{Results}\label{sec:experiments}

Offline reinforcement learning evaluation is conducted in a terrain-aware simulator for UAV-assisted relaying that  captures UAV mobility, user motion, and terrain-induced wireless feasibility constraints. Learning is performed strictly offline using a fixed dataset collected from heterogeneous behavior policies. All methods are trained and evaluated under identical conditions and differ only in how the terrain-context observation is represented. Key environment, dataset, and training parameters used with CQL are summarized in Table~\ref{tab:setup}. Performance is reported as averages over three random seeds for training and evaluation.

\begin{table}[ht]
\centering
\caption{Environment, dataset, representation, and training parameters.}
\label{tab:setup}
\small
\begin{tabular}{p{0.38\columnwidth} p{0.55\columnwidth}}
\toprule
\textbf{Parameter} & \textbf{Value} \\
\midrule
\multicolumn{2}{l}{\textbf{Terrain / Map}} \\
Map type & Digital elevation + land cover \\
Map resolution & 400 m \\
Land-cover model & Discrete classes (pathloss offsets) \\
\midrule
\multicolumn{2}{l}{\textbf{Users}} \\
Number of users $U$ & 3 \\
Mobility model & Waypoint-based \\
User velocity & 10 m/s \\
User height & 1.5 m \\
\midrule
\multicolumn{2}{l}{\textbf{UAV / Network}} \\
Number of UAVs & 1 \\
Base station height & 20 m \\
Action dimension & 3 \\
Episode length $T$ & 600 steps \\
\midrule
\multicolumn{2}{l}{\textbf{State / Dataset}} \\
State dimension & 5136 \\
State components & Terrain map + kinematics + link features \\
Dataset size $N$ & 320k transitions \\
Behavior policies & Random, centroid, coverage-oriented \\
Oracle-assisted data & Enabled (dataset only) \\
\midrule
\multicolumn{2}{l}{\textbf{Representation}} \\
Representations & Raw, VAE, AE, PCA \\
Latent dimensions $d_z$ & 32, 64, 128 \\
\midrule
\multicolumn{2}{l}{\textbf{CQL Training}} \\
Batch size & 128 \\
Discount factor $\gamma$ & 0.99 \\
Training steps & $1\times10^6$ \\
Q learning rate & $3\times10^{-5}$ \\
Actor learning rate & $5\times10^{-5}$ \\
CQL coefficient $\alpha$ & 0.5 \\
Target update rate $\tau$ & 0.005 \\
\bottomrule
\end{tabular}
\end{table}

\subsection{Methods Compared}
Several variants of Conservative Q-Learning (CQL) that differ only in
their state representation are compared. \textbf{CQL (Raw)} operates directly on the full high-dimensional observation, consisting of a flattened terrain map concatenated with kinematic and link-quality features. To study representation learning, \textbf{CQL+VAE} is evaluated, where observations are first compressed into a low-dimensional latent space using a variational autoencoder. Latent dimensions $d_z \in \{32,64,128\}$ are considered, with $d_z=64$ used unless otherwise stated. Additionally, \textbf{CQL+AE} uses an autoencoder to learn a nonlinear low-dimensional embedding without variational regularization~\cite{Hinton2006AE}, while \textbf{CQL+PCA} applies principal component analysis, a linear dimensionality-reduction method based on maximum-variance projections~\cite{Jolliffe2002}.
 All methods share identical policy and value network architectures and CQL hyperparameters, ensuring that observed differences arise solely from the state representation.

\subsection{Evaluation Metrics}
Performance is evaluated using metrics that capture both realized relay service and physical feasibility. The average and peak numbers of users served per episode are reported, as well as performance normalized by the candidate-set feasibility upper bound (CS-FUB). To quantify control suboptimality, the distribution of CS-FUB optimality gaps is analyzed.
 Finally, learning efficiency is assessed using the \emph{time-to-feasible} metric, defined as the number of steps required to first achieve feasible service when the CS-FUB indicates that such service is physically achievable.

\subsection{Main Results}

\begin{figure}[ht]
    \centering
    \includegraphics[width=\linewidth]{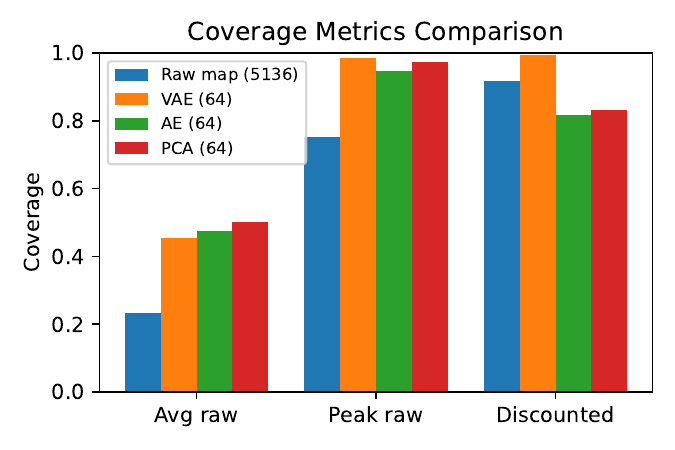}
    \caption{Comparison of average served users, peak served users, and discounted (CS-FUB-normalized) performance for latent and raw-state policies.}
    \label{fig:coverage_metrics}
\end{figure}

\paragraph{Raw terrain observations hinder offline learning}
Fig.~\ref{fig:coverage_metrics} presents the coverage performance of CQL (Raw), CQL+AE, CQL+PCA, and CQL+VAE. Notably, \textbf{CQL (Raw)} achieves the lowest average and peak coverage, despite having access to the richest state information.

\begin{figure}[t]
    \centering
    \includegraphics[width=\linewidth]{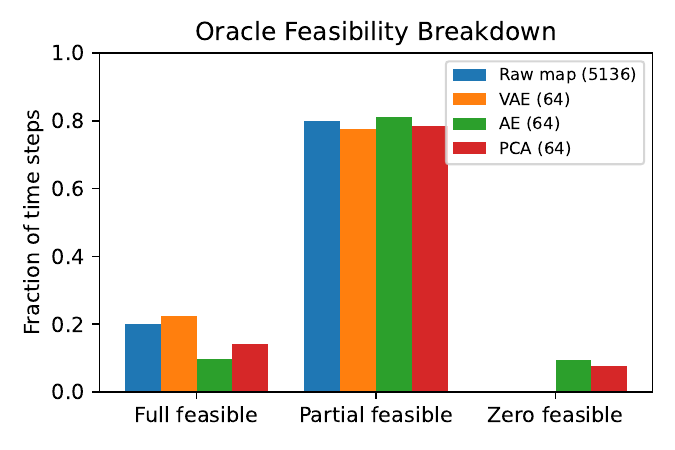}
    \caption{CS-FUB feasibility breakdown showing the fraction of time steps where full, partial, or no user service is achievable under the candidate-set feasibility upper bound.}
    \label{fig:csfub_feasibility}
\end{figure}

\begin{figure}[t]
    \centering
    \includegraphics[width=\linewidth]{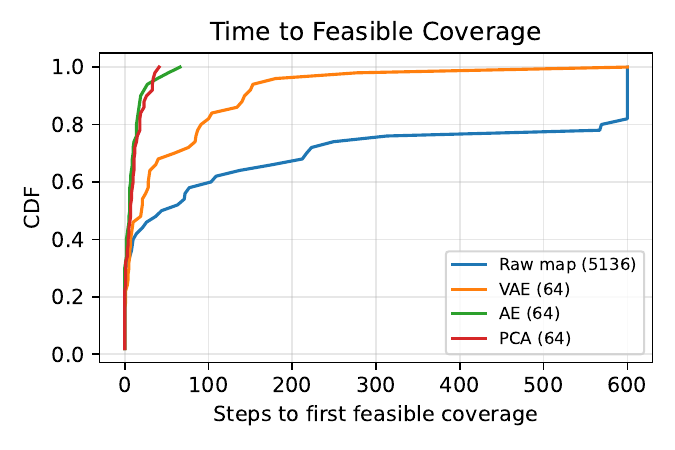}
    \caption{CDF of time steps required to reach a feasible target service level (when CS-FUB indicates feasibility). Latent policies reach feasible service significantly faster.}
    \label{fig:time_feasible}
\end{figure}

\paragraph{Compact representations recover feasible performance}
Latent representations dramatically improve both absolute and normalized performance. VAE- and PCA-based policies recover a significantly larger fraction of the CS-FUB and achieve consistently higher realized coverage
(Fig.~\ref{fig:coverage_metrics}). Among all methods, \textbf{CQL+VAE} yields the strongest overall performance across evaluated metrics.

\paragraph{Feasibility exists but is not exploitable from raw states.}
The feasibility breakdown in Fig.~\ref{fig:csfub_feasibility} shows that raw terrain observations admit a large fraction of fully or partially feasible time steps under the CS-FUB. However, this theoretical feasibility is not translated into realized service due to learning instability in high-dimensional state spaces. In contrast, compact representations slightly reduce achievable performance relative to the CS-FUB but enable policies to exploit feasible configurations more reliably.

\paragraph{Compact representations improves learning efficiency}
Fig.~\ref{fig:time_feasible} shows the cumulative distribution of time to first
feasible service. Latent-state policies reach feasible relay configurations
significantly faster than raw-state policies, indicating
improved learning stability and sample efficiency under offline training.

\subsection{Ablation Studies}
Several ablation studies were conducted to isolate the effect of representation design.
First, varying the latent dimension reveals that moderate compression
($d_z=64$) provides the best tradeoff between performance and stability; excessive compression discards task-relevant information, while insufficient compression limits the benefits of dimensionality reduction. Second, comparing VAE, AE, and PCA at comparable latent sizes shows that all compact representations outperform raw-state CQL. VAE achieves the strongest overall performance, while PCA remains competitive despite its simplicity. However, AE and PCA can reach feasible configurations faster, indicating a trade-off between learning speed and final performance.

\section{Conclusion}\label{sec:conclusion}
In this paper, offline reinforcement learning for UAV-assisted relaying in large-scale outdoor environments with wireless conditions affected by the terrain, where the agent observes high-dimensional map-like contextual inputs are explored and optimized. The CS-FUB is introduced to explicitly characterize fundamental access--backhaul feasibility limits and to distinguish unavoidable physical limits from control sub-optimality. To address the challenges  of learning  from high-dimensional terrain context under fixed offline datasets, compact state representations obtained via VAE compression are explored  in combination with CQL. Simulation results demonstrate that directly applying offline RL to raw high-dimensional terrain-context observations leads to slow convergence and large feasibility gaps, despite access to rich information. In contrast, VAE-encoded latent representations substantially improve learning stability and enable earlier convergence to feasible relay configurations, reducing control sub-optimality relative to physical limits.
Future work will consider extensions to multi-UAV relay systems, temporal representation learning, adaptive candidate-set construction for tighter feasibility bounds, and uncertainty-aware latent representations to improve robustness under sparse or biased offline data.

\vspace{-3mm}
\section*{Acknowledgment}
This publication has emanated from research supported in part by grants from Research Ireland under CONNECT grant number \text{13/RC/2077\_P2} and the Royal Society University Research Fellowship 24/RS-URF/3981. For the purpose of Open Access, the author has applied a CC BY public copyright licence to any Author Accepted Manuscript version arising from this submission.

\end{document}